\documentclass[5p]{elsarticle}
\biboptions{sort&compress}   
\usepackage{hyperref}
\usepackage{amsmath}
\usepackage{amssymb}
\usepackage{flushend}
\journal{Elsevier}
\newcommand{\be}{\begin{equation}}
\newcommand{\ee}{\end{equation}}
\newcommand{\ba}{\begin{eqnarray}}
\newcommand{\ea}{\end{eqnarray}}
\begin{document}
\begin{frontmatter}
\title{The scalar glueball operator, the $a$-theorem, and the onset of conformality}
\author[rug]{T. Nunes da Silva\fnref{addressfootnote}}
\ead{t.j.nunes@posgrad.ufsc.br}
\fntext[addressfootnote]{Present address: Departamento de F\'isica, CFM, Universidade Federal de Santa Catarina, 88040-900, Florian\'opolis, Brazil.}
\author[rug]{E. Pallante\corref{cor1}}
\ead{e.pallante@rug.nl}
\author[rug]{L. Robroek}
\ead{l.l.robroek@student.rug.nl}
\cortext[cor1]{Corresponding author}
\address[rug]{Van Swinderen Institute for Particle Physics and Gravity, Nijenborgh 4, 9747 AG, Groningen, The Netherlands}
\begin{abstract}
We show that the anomalous dimension $\gamma_G$ of the scalar glueball operator contains information on the mechanism that leads to the onset of conformality at the lower edge of the conformal window in a non-Abelian gauge theory. In particular, it distinguishes whether the merging of an UV and an IR fixed point -- the simplest mechanism associated to a conformal phase transition and preconformal scaling -- does or does not occur. 
At the same time, we shed light on new analogies between QCD and its supersymmetric version.
In SQCD, we derive an exact relation between $\gamma_G$ and the mass anomalous dimension $\gamma_m$, and we prove that the SQCD exact beta function is incompatible with merging as a consequence of the $a$-theorem; we also derive the general conditions that the latter imposes on the existence of fixed points, and prove the absence of an UV fixed point at nonzero coupling above the conformal window of SQCD.   
Perhaps not surprisingly, we then show that an exact relation between $\gamma_G$ and $\gamma_m$, fully analogous to SQCD, holds for the massless Veneziano limit of large-N QCD. We argue, based on the latter relation, the $a$-theorem, perturbation theory and physical arguments, that the incompatibility with merging may extend to QCD. 
\end{abstract}
\begin{keyword}
Non-Abelian gauge theories\sep QCD\sep conformal symmetry\sep conformal window
\end{keyword}
\end{frontmatter}
\section{Introduction}
For a sufficiently large number $N_f$ of mass\-less fer\-mi\-ons, it is believed that a new phase of QCD arises \cite{Caswell1974,Banks:1981nn}. It is called the conformal window, ranging from a value $N_f^c$, where the zero-temperature theory deconfines and chiral symmetry is restored, to a value $N_f^{AF}$, above which asymptotic freedom is lost. Theories with $N_f^c<N_f<N_f^{AF}$ have a nontrivial, i.e.,  interacting infrared (IR) fixed point where they are conformal. A conformal window also arises in supersymmetric versions of non-Abelian gauge theories \cite{Seiberg:1994pq} and generalisations of QCD with fermions in higher dimensional representations and/or other gauge groups. Theories with $N_f>N_f^c$ may thus lead to new possibilities for particle dynamics. 
Above the conformal window, $N_f>N_f^{AF}$, infrared freedom leads to the possibility of realising ``asymptotically safe'' theories, with a nontrivial ultraviolet (UV) fixed point, see e.g. \cite{Litim:2014uca}. 

Just below the conformal window, $N_f\lesssim N_f^c$, it has been proposed the phenomenologically interesting possibility of a preconformal behaviour characterised  by a walking, i.e., slow-running \footnote{At least on a finite energy range $[\mu_{IR}, \mu_{UV}]$.} gauge coupling \cite{1986PhRvL..57..957A, 1991PhRvD..44.2678L}. Since theories with a preconformal behaviour would not differ from QCD as far as their fixed point structure is concerned, they must be confining and asymptotically free\footnote{In other words, no phase transition is expected to occur between QCD and preconformal theories with $N_f\lesssim N_f^c$ at zero temperature.}. However, the preconformal behaviour  is entangled to the nature of the mechanism that opens the conformal window at $N_f^c$, and it should be expected to modify the evolution from the UV to the IR of observables.  It has been shown that a phase transition named conformal in \cite{PhysRevD.58.105017, Miransky:1996pd, Miransky1985} -- the equivalent of a Berezinskii-Kosterlitz-Thouless (BKT) phase transition in two-dimen\-sional spin systems \cite{Berezinsky:1970fr, Berezinsky:1972, Kosterlitz:1973xp} -- leads to the walking phenome\-non for $N_f\lesssim N_f^c$, and the associated preconformal behaviour of physical observables known as Miransky or BKT scaling \cite{PhysRevD.58.105017, Miransky:1996pd, Miransky1985, Berezinsky:1970fr, Berezinsky:1972, Kosterlitz:1973xp}. Interestingly, it was then observed \cite{Kaplan:2009kr} that the merging of a pair of UV and IR fixed points at $N_f^c$  is a simple way of realising preconformal scaling. Alternatively, and among other possibilities, a first order phase transition at $N_f^c$ would not lead to precursor effects, see \cite{Sannino:2012wy}  in this context. 

It is thus relevant to identify observables that carry the imprint of the mechanism for the onset of conformality at $N_f^c$, and at the same time are stringently constrained by universal principles, such as exact symmetries and the ultraviolet to infrared renormalisation group (RG) flow governed by the $a$-theorem.  

In this letter we show that the anomalous dimension $\gamma_G$ of the scalar glueball operator at a fixed point is such an observable; its ultraviolet to infrared flow determines whether an UV-IR fixed point merging occurs. We also show that for both SQCD, 
for which an exact beta function is known \cite{Novikov1983381, Novikov1986329, Shifman1986456},
and the massless Veneziano limit of large-N QCD, whose exact beta function has been recently proposed \cite{Bochicchio:2008vt,Bochicchio:2013aha,Bochicchio:2012bj}, there is an exact relation between $\gamma_G$ and the mass anomalous dimension $\gamma_m$, thus relating the two RG flows. These are in turn governed by the $a$-theorem, which allows us to prove the incompatibility of SQCD with merging to all orders in perturbation theory, and directly constrain the existence of an UV fixed point at nonzero coupling. The analogies with SQCD and the universality of the $a$-theorem suggest that the same incompatibility may extend to QCD. 
Indeed, though the exact beta function proposed in \cite{Bochicchio:2013aha} for Veneziano large-N QCD has been obtained by means 
of homology methods \cite{Bochicchio:2008vt,Bochicchio:2013aha,Bochicchio:2012bj} that are not as much consolidated in quantum field theory as their cohomological counterparts involving supersymmetry, it passes a number of perturbative and nonperturbative consistency checks, as we will discuss in section \ref{sec:largeN_results}.  

The letter is organised as follows. In section \ref{sec:gammaG} we review a known formula for $\gamma_G$ based on the trace anomaly.
In section \ref{sec:pert_results} we analyse $\gamma_G$ in two-loop perturbation theory and close to the upper edge $N_f\lesssim N_f^{AF}$, partly reviewing known results, and we comment on the limits of applicability of perturbation theory in this context. In section \ref{sec:SQCD_results} we derive results in SQCD, and prove the incompatibility with merging in \ref{sec:UV-IRmerging}.
In section \ref{sec:largeN_results} we discuss the massless Veneziano limit of large-N QCD, and investigate to what extent it reproduces the results of SQCD.
As a side note, in section \ref{sec:merge_multiple} we discuss why the addition of effective four-fermion operators does not lead to alternative viable realisations of merging in QCD. 
We conclude in section \ref{sec:conc}.
\section{The scalar glueball operator and its anomalous dimension}
\label{sec:gammaG}
It is well known that the anomalous dimension of the scalar glueball operator $\mbox{Tr} (G^2)\equiv G_{\mu\nu}^a G^{a \mu\nu}$ is constrained by the trace anomaly, i.e., the nonzero contribution to the trace of the energy-momentum tensor, see, e.g., \cite{Nielsen:1977sy} and more recently \cite{Bochicchio:2013tfa, Crewther:2013vea}. The trace anomaly of QCD that enters the matrix elements of renormalised gauge invariant operators is\footnote{We are thus not interested in the most general expression, which also involves gauge-fixing and EoM operators, see \cite{Nielsen:1977sy,Collins1977,Adler1977}.} 
\be
\label{eq:tracea}
T^\mu_\mu = \frac{\beta (g )}{2g}\,\mbox{Tr} (G^2) + {\mbox{fermion mass contribution}}\, ,
\ee
with the beta function $\beta (g )\!=\!{\partial g (\mu )}/{\partial\log\mu}$ for given $N$ colours and $N_f$ flavours; an analogous relation is valid in SQCD. We shall restrict ourselves here to the massless theory.
The nonrenormalisation of $T^\mu_\mu$ implies that the renormalised operator $O_{\text{RGI}}\!\equiv\!(\beta (g )/g)\mbox{Tr} (G^2)$ is also renormalisation-group (RG) invariant, i.e., $d O_{\text{RGI}}$ $/d\log\mu\!=\!0$.
Using inside the latter equation a Callan-Symanzik equation for the renormalised operator $\mbox{Tr} (G^2)$
\be
 \frac{d}{d\log\mu}\left ( Z^{-1}_{G} \mbox{Tr} (G^2)\right ) =0\, ,~~~~\gamma_G(g) = -\frac{\partial \log Z_{G}}{\partial\log\mu}\, ,
\ee
with $\gamma_G(g)$ the anomalous dimension  of $\mbox{Tr} (G^2)$ for $N$ and $N_f$ fixed, one obtains 
\be
\frac{d}{d\log\mu}\left ( \frac{\beta (g )}{g} Z_{G}\right )=0 
\ee
and
\be
\label{eq:gammaG}
\gamma_G(g) = g\frac{\partial}{\partial g}\left (\frac{\beta (g )}{g}   \right )\,=\,\beta ' (g ) -\frac{\beta (g )}{g}
\ee
 for a theory with given $N$ and $N_f$. 
This equation reproduces the known result in perturbative QCD \cite{Kataev:1981gr, Chetyrkin:1997iv, Bochicchio:2013tfa},  $\gamma_G=-2\beta_0 g^2 +\ldots$, $\beta_0$ from (\ref{eq:b12}), and $\gamma_G$ is negative, so  that 
the operator $\mbox{Tr}(G^2)$ becomes increasingly relevant towards the infrared. 

We shall be interested in the $g$ and $N_f$ dependence of the anomalous dimensions, thus in general $\gamma_G(g,N_f)$. At a fixed point of the renormalization group flow, the solution of $\beta(g,N_f)\!=\!0$ thus defines the function $g^*(N_f)$ of fixed-point couplings on the plane $(g,N_f)$, and equation (\ref{eq:gammaG}) provides $\gamma_G$ at $g^*(N_f)$:
\be
\label{eq:gammaG_FP}
\gamma_G^* (N_f)\equiv \gamma_G(g,N_f)|_{g\!=\!g_*(N_f)}= \beta' (g, N_f)|_{g\!=\!g_*(N_f)}\, ,
\ee
where the prime will always denote the derivative with respect to $g$. The fixed-point anomalous dimension $\gamma_G^*$ for a given $N_f$
is a physical property of the system, renormalisation scheme independent; the scaling dimension of $\mbox{Tr} (G^2)$, $d_G=4+\gamma_G^*$, thus enters the exact conformal scaling of the corresponding correlators at the fixed point. 
\section{Perturbative Results in QCD}
\label{sec:pert_results}
It is instructive to first recall some features of perturbation theory.
The QCD beta function 
can be expressed as a series
\begin{equation}
\beta (g ) = -g^3\sum_{l=0}^\infty{\beta_l\, g^{2l}}\, , 
\label{eqn:beta-expansion}
\end{equation}
where $(l+1)$ denotes the number of loops involved in the calculation of $\beta_l$.
The coefficients $\beta_{0,1}$ are universal \cite{Gross:1973id, PhysRevLett.30.1346, Caswell1974, Jones1974531}, i.e., renormalisation scheme independent, given by
\begin{equation}
\label{eq:b12}
\begin{gathered}
\beta_0 = \frac{1}{3(4\pi)^2}(11 C_A - 4T_f N_f)\\
\beta_1 = \frac{1}{3(4\pi)^4}\left[ 34C_A^2 - 4(5C_A + 3C_f)T_fN_f\right],
\end{gathered}
\end{equation}
here written in terms of the quadratic Casimir invariants $C_f \equiv C_2(R)$ and $C_A \equiv C_2(G)$,  for, respectively, the representation $R$ to which the $N_f$ fermions belong and the adjoint representation. The quantity $T_f \equiv T(R)$ is the trace invariant for the representation $R$.
Coefficients of higher order are renormalisation scheme dependent \cite{PhysRevD.8.3633, GrossLesHouches} and have been calculated up to five-loop order in the $\overline{MS}$ scheme
 \cite{Tarasov:1980au, Larin:1993tp, vanRitbergen:1997va,Baikov:2016tgj}. 

To two loops, a nontrivial IR fixed point with coupling $g_*^2\!=\!-{\beta_0}/{\beta_1}$ is one root of the equation 
$\beta(g)\!=\!0$ for some given $N_f$, 
and from (\ref{eq:gammaG_FP}) $\gamma_G^*\!=\!-2{\beta_0^2}/{\beta_1}$. 

We are interested in the way $\gamma_G^*$ varies along the curve of IR fixed points $g_*(N_f)$ as $N_f$ decreases in the conformal window of QCD, i.e., for $N_f$ Dirac fermions in the fundamental representation; in this case, 
$N_f^{AF}\!=\!(11/2)N$, $C_A\!=\!N$, $C_f\!=\!(N^2-1)/(2N)$, and $T_f\!=\!1/2$ in (\ref{eq:b12}). 
In the Veneziano limit, $N,\,N_f\rightarrow\infty$, holding $x\!=\!N_f/N$ and $Ng^2$ constant, and $\epsilon\!=\! 11/2 - N_f/N\ll 1$, that is close to the upper edge, one obtains 
$Ng_*^2/(4\pi )^2\!\simeq\! (4\epsilon /75) (1+O(\epsilon ))$ \cite{Banks:1981nn}, and $\gamma_G^*\!\simeq\! (16\epsilon^2/225) (1+O(\epsilon ))$ positive \footnote{For a study of higher orders in perturbation theory see \cite{Shrock:2013pya, Shrock:2013ca}.}. 
Its derivative with respect to $N_f\!=\!xN$ with fixed $N$ and $x$ continuous in the Veneziano limit  
\be
\label{eq:pQCDgamma}
\frac{d\gamma_G^*}{dN_f} = -\frac{32\epsilon}{225 N} (1+O(\epsilon ))
\ee
is negative and of order $\epsilon /N$, thus implying that $\gamma_G^*$, as $Ng_*^2$, is a strictly monotonic function of $N_f$ along the IR fixed point curve, at least in the neighbourhood of the upper edge, and it increases as $N_f$ decreases. 
In other words, the universal two-loop contribution in perturbation theory is consistent with an increasingly irrelevant operator $\mbox{Tr}(G^2)$ as approaching the lower edge. 

We finally observe that, beyond the Veneziano limit and moving away from the upper edge, the two-loop expression $\gamma_G^*\!=\! -2{\beta_0^2}/{\beta_1}$ remains indeed positive and monotonically increasing as $N_f$ decreases on the entire interval $ N_f^c \leqslant N_f\leqslant N_f^{AF}$, and the IR zero disappears at $N_f^c$ due to the change of sign of $\beta_1$;  for $N\!=\!3$ and $N_f$ fundamental  fermions, this occurs at $N_f\!\sim\! 8.05$. 
The change of sign of $\beta_1$ would imply that the fixed point disappears at infinite coupling $g_*^2\!=\!-{\beta_0}/{\beta_1}\rightarrow\infty$, and the same singularity occurs for $\gamma_G^*$. This behaviour, however, is likely to be an artefact of the truncated perturbative expansion, as we further discuss  in section \ref{sec:largeN_results}. Also, since the perturbative series (\ref{eqn:beta-expansion}) is at best asymptotic, we should take the two-loop, or higher order, results at most as qualitative indications. 
\section{Results in SQCD}
\label{sec:SQCD_results}
We consider $SU(N)$ supersymmetric QCD (SQCD) with $N_f$ fundamental flavours $Q^i$ in the $N$ representation and $\tilde{Q}_{\tilde{i}}$ in the $\bar{N}$ representation ($i,\tilde{i}\!=\!1,\ldots N_f$), where many results can be derived exactly. Our goal in this section is to determine exact constraints on the UV to IR flow of $\gamma_G(g,N_f)$ and the mass anomalous dimension $\gamma_m(g,N_f)$. Later on, in section \ref{sec:largeN_results}, we will find that some properties of $\gamma_G$ can be proved to be equally true in the massless Veneziano limit of large-N QCD.  
For our purpose, we make use of Seiberg's solution for the phases of SQCD \cite{Seiberg:1994pq}, the NSVZ exact beta function \cite{Novikov1983381, Novikov1986329, Shifman1986456}, and the $a$-theorem on the irreversibility of renormalisation group (RG) flows in four-dimensional field theories \cite{Komargodski:2011vj}, a generalisation to higher dimensions of Zamolodchikov's $c$-theorem \cite{Zamolodchikov:1986gt} in two dimensions. 
\subsection{Known results in SQCD}
\label{sec:knownSQCD}
The NSVZ exact beta function for given $N$ and $N_f$ reads \cite{Novikov1983381, Novikov1986329, Shifman1986456} 
\be
\label{eq:NSVZ}
\beta (g)=-\frac{g^3}{16\pi^2}\, \frac{3N-N_f +N_f\gamma_m(g)   }{ 1-Ng^2/(8\pi^2)}\, ,
\ee
with 
\be
\label{eq:SQCDgamma_m}
\gamma_m (g) = -\frac{g^2}{8\pi^2}\frac{N^2-1}{N} + O(g^4)\, 
\ee
the mass anomalous dimension computed in perturbation theory. A powerful property of SQCD is that its exact beta function and the global anomaly free R symmetry at a fixed point determine exactly the mass anomalous dimension  $\gamma_m^*(N_f)$ along the curve of IR fixed points, $g_*(N_f)$, in the conformal window of SQCD, which extends on the interval $3N/2<N_f<3N$ \cite{Seiberg:1994pq}. 

In more detail, the exact R symmetry at a fixed point allows us to determine the anomalous dimension of spinless chiral primary operators from their R-charge. For the gauge invariant composite meson operator $M\!=\!\tilde{Q}Q$, with scale dimension $D_{\tilde{Q}Q}$ and R-charge $R_{\tilde{Q}Q}$, one has \cite{Seiberg:1994pq}
\be
\label{eq:QQ}
D_{\tilde{Q}Q}=\frac{3}{2}R_{\tilde{Q}Q}=3R= 3\frac{N_f-N}{N_f}\,,
\ee
with $R$ the R-charge of $Q$($\tilde{Q}$), and the last equality dictated by the R-charge assignments of $Q$($\tilde{Q}$) under $U(1)_R$. Using $D_{\tilde{Q}Q}\!=\!2+\gamma_m^*$, one obtains $\gamma_m^*$ exactly
\be
\label{eq:gamma_at_IRFP}
\gamma_m^*(N_f) = 1-\frac{3N}{N_f}\, ,
\ee
which is indeed a zero of the beta function (\ref{eq:NSVZ}), 
provided the pole is not hit, i.e., $Ng_*^2/(8\pi^2)<1$ \footnote{Note, however, that the cusp singularity in (\ref{eq:NSVZ}) for $Ng^2/(8\pi^2)\!=\!1$ is a renormalisation-scheme dependent condition; it cannot occur if a physical zero of the numerator of (\ref{eq:NSVZ}) occurs. The role of the cusp singularity in SUSY Yang-Mills $(N_f\!=\!0)$, where (\ref{eq:NSVZ}) has no zero for $g>0$, is discussed in \cite{Kogan:1995hn}.}. 
Equation (\ref{eq:gamma_at_IRFP}) is then taken to determine $\gamma_m$ along the curve of IR fixed points in the conformal window with varying $N_f$; indeed it vanishes at the upper edge, $N_f\!=\!3N$, where the theory is IR free, and it is negative below it.

The lower edge is signaled by a physical condition, i.e., a renormalisation-scheme independent condition. This is the saturation of the unitarity bound in Seiberg's solution for the phases of SQCD, and such a condition is independent of the beta function. Specifically, the saturation of the unitarity bound $D_{\tilde{Q}Q}\!=\!1$ implies $\gamma_m^*\!=\!-1$, which in turn implies that the numerator of the beta function has a zero for $N_f\!=\!3N/2$. This identifies the lower edge of the conformal window for SQCD. 

Equation (\ref{eq:gamma_at_IRFP}) is implicitly a function of the coupling $g^*(N_f)$ along the IR fixed point curve. 
One can determine $g^*(N_f)$ perturbatively, by taking $N, N_f\to \infty$ and holding $Ng^2$ and $x\!=\!N_f/N$ constant, with $\epsilon\!=\! 3-N_f/N\ll 1$, i.e., close to the upper edge of the conformal window. This gives  \cite{Seiberg:1994pq}
\be
\label{eq:gIRFP}
Ng_*^2 = \frac{8}{3}\pi^2\epsilon+O(\epsilon^2)\,.
\ee 
\subsection{New results in SQCD}
\label{sec:newSQCD}
Equation (\ref{eq:gammaG_FP}) implies that 
the derivative of (\ref{eq:NSVZ}) with respect to the coupling, evaluated at a nontrivial fixed point,  gives the anomalous dimension of the scalar glueball operator at the fixed point as a function of $N_f$
\be
\label{eq:NSVZderiv}
\gamma_G^*(N_f)=-\frac{g_*^3}{16\pi^2}\,\frac{N_f\gamma_m^{\prime *}(N_f)   }{ 1-Ng_*^2/(8\pi^2)}\, ,
\ee
 where analogously to (\ref{eq:gammaG_FP}) $\gamma_m^{\prime *}(N_f)\equiv \gamma'_m(g,N_f)|_{g\!=\!g_*(N_f)}$ and 
the prime denotes the derivative with respect to $g$. Equation (\ref{eq:NSVZderiv}) establishes a useful relation between the anomalous dimension $\gamma_G$ and the derivative of $\gamma_m$. This is a key result that we are going to use  in the rest of this section. In particular, our task is to derive constraints on the flow of $\gamma_m$, its derivatives, 
 and $\gamma_G$, using equation (\ref{eq:NSVZderiv}), R symmetry and the $a$-theorem.

It is convenient to immediately summarise the main new results of section \ref{sec:SQCD_results} for SQCD. They are all valid in the conformal window and its edges, $3N/2\leqslant N_f\leqslant 3N$, and can be summarised as follows:
\begin{itemize}
\item[1)] $\gamma_m(g,N_f)$ is a strictly monotonic function of $g$ for $N_f$ fixed, for any valid RG flow from an UV fixed point to an IR fixed point, and it may be stationary at the fixed points. This result is implied by the $a$-theorem and proved in section \ref{sec:atheorem}, equation (\ref{eq:a_vs_g}). The strict monotonicity of $\gamma_m$ away from the fixed point will be sufficient to prove the incompatibility of the SQCD exact beta function (\ref{eq:NSVZ}) with merging in section \ref{sec:UV-IRmerging}.
\item[2)] A result stronger than the incompatibility with merging is also  proved in section \ref{sec:UV-IRmerging}: In SQCD, 
the $a$-theorem implies through equation (\ref{eq:a_vs_g}) that the beta function, if continuous and thus free from cusp singularities, does not admit more than one fixed point at nonzero coupling. Hence, in the conformal window and its lower edge, $3N/2\leqslant N_f<3N$, the existence of the IR fixed point at nonzero coupling excludes an UV fixed point at nonzero coupling.
If instead one of the two fixed points occurs at zero coupling, the $a$-theorem can be satisfied, but not always. We find under which conditions  the $a$-theorem is satisfied.
\end{itemize}
Results 1) and 2) are nevertheless not able to determine if $\gamma_G^*(N_f)$ is strictly positive along the nontrivial  IR fixed point curve $g_*(N_f)$ of SQCD, or it vanishes. Results 3) to 5) below provide arguments in favour of a strictly positive $\gamma_G^*(N_f)$  for $3N/2\leqslant N_f< 3N$.
\begin{itemize}
\item[3)]  We know exactly $\gamma_m^*(N_f)$ along the IR fixed point curve. If $N_f$ is assumed to be continuous, then  
$d\gamma_m^*/$ $dN_f >0$, and thus $\gamma_m^*(N_f)$ is strictly monotonic in $N_f$ and decreases as $N_f$ decreases along $g_*(N_f)$. Also, $d^2\gamma_m^*/dN_f^2<0$ implies that  $d\gamma_m^*/dN_f$ itself strictly increases as $N_f$ decreases. This result comes straightforwardly from the exact solution for $\gamma_m^*(N_f)$ in SQCD and the $a$-theorem.
\item[4)] In the Veneziano limit, to leading order in perturbation theory and close to the upper edge, the IR fixed point coupling $Ng_*^2$ is strictly monotonic in $x\!=\!N_f/N$, and, with abuse of notation, in  
 $N_f\!=\!xN$ with fixed $N$ and $x$ continuous in the Veneziano limit. 
This result is fully analogous to the perturbative QCD result in section \ref{sec:pert_results}. 
\item[5)] In the Veneziano limit, to leading order in perturbation theory and close to the upper edge, the solution for the IR fixed point of SQCD is consistent with $\gamma_m^{\prime *}(N_f)\!<\!0$, and, through (\ref{eq:NSVZderiv}),
a strictly positive $\gamma_G^*(N_f)$. 
We add that a result fully analogous to that of QCD two-loop perturbation theory in (\ref{eq:pQCDgamma}), is obtained in SQCD if the two-loop SQCD beta function is used.  
\end{itemize}
Result 3) is straightforwardly implied by taking the derivatives of (\ref{eq:gamma_at_IRFP}), specifically, $d\gamma_m^*/dN_f\!=\! 3N/N_f^2$ and $d^2\gamma_m^*/dN_f^2$ $\!=\! -6N/N_f^3$.

The derivative of (\ref{eq:gIRFP}) with respect to $N_f\!=\!xN$ for $N$ fixed, $\partial (Ng_*^2)/\partial N_f\!=\!-8\pi^2/3N (1+O(\epsilon))$ is negative to leading order, so is $\partial g_*/\partial N_f$, thus providing result 4); it agrees with the observation that the theory is increasingly strongly coupled as $N_f$ decreases. 

Result 5) follows from (\ref{eq:NSVZderiv}) and the properties of $\gamma_m$.
In fact, the derivative with respect to $N_f$
\be
\label{eq:gammaflow}
\frac{d\gamma_m^*}{dN_f} = \left.{\frac{\partial\gamma_m(g,N_f)}{\partial N_f} }\right|_{g\!=\!g_*(N_f)} + \gamma_m^{\prime *}(N_f) 
\left (\frac{\partial g_*}{\partial N_f}\right )
\ee
is known exactly, $d\gamma_m^*/dN_f\!=\!3N/N_f^2$. 

The rhs of (\ref{eq:gammaflow}) can be determined in the Veneziano limit with $\epsilon\ll 1$, and taking derivatives with respect to $N_f\!=\!xN$ for $N$ fixed. Equation 
(\ref{eq:SQCDgamma_m})  gives $\partial\gamma_m/\partial N_f\!=\!0$ to leading order and, using (\ref{eq:gIRFP}), the expansion
\be
\gamma_m^{\prime *}(N_f) 
 \left(\frac{\partial g_*}{\partial N_f}\right ) = \frac{1}{3N}\left (1-\frac{1}{N^2}\right ) (1+O(\epsilon))\, 
\ee
reproduces the expansion $d\gamma_m^*/dN_f\!=\!3N/N_f^2\!=\!1/(3N) (1+O(\epsilon))$ to the leading $1/N$ order.
This result is consistent with $\gamma_m^{\prime *}(N_f)\!<\!0$ and, through (\ref{eq:NSVZderiv}),
$\gamma_G^*(N_f)$ strictly positive. 
\subsection{Implications of the $a$-theorem}
\label{sec:atheorem}
The $a$-theorem for four-dimensional RG flows establishes the existence of a monotonically decreasing function that interpolates between the Euler anomalies of an UV and an IR CFT, i.e., $a_{UV}-a_{IR}>0$. This function also provides an effective measure of the number of massless degrees of freedom, consistently with the intuition that this number decreases as we integrate out high momenta. 
Cardy's conjectured $a$-function \cite{Cardy:1988cwa}, given by the integral of the trace of the energy-momentum tensor on the sphere $S^4$, has passed all tests in the context of theories that are free in the UV and whose IR dynamics can be computed. The recent proof  of the $a$-theorem \cite{Cardy:1988cwa, Komargodski:2011vj} requires the rather general prerequisite of a unitary S matrix.  

We use here the interpolating $a$-function for SQCD in the conformal window obtained in \cite{Anselmi:1997ys}, whose IR value can be computed from the $U(1)_R FF$, $U(1)_R$ and $U(1)_R^3$ anomalies, and provides $a_{UV}-a_{IR}$ in terms of the anomaly free R-charge of the field $Q$($\tilde{Q}$). The latter is a function of $\gamma_m$ via (\ref{eq:QQ}) and $D_{\tilde{Q}Q}\!=\!2+\gamma_m$.
This means that the $a$-theorem directly constrains the UV to IR flow of $\gamma_m(g,N_f)$ in SQCD, and, via
 (\ref{eq:NSVZderiv}), it constrains that of $\gamma_G(g,N_f)$. 

Without knowledge of the $a$-theorem, equation (\ref{eq:gamma_at_IRFP}) already implies that, along the IR fixed point curve, $D_{\tilde{Q}Q}$, $R$ and $\gamma_m^*$ decrease from their value at the upper edge ($D_{\tilde{Q}Q}\!=\!2$, $R\!=\!2/3$, $\gamma_m^*\!=\!0$) to their value at the lower edge ($D_{\tilde{Q}Q}\!=\!1$, $R\!=\!1/3$, $\gamma_m^*\!=\!-1$), where the unitarity bound is saturated. 

The $a$-theorem allows us to further establish the monotonic variation of $\gamma_m(g,N_f)$ along any valid RG trajectory from the ultraviolet to the infrared. In particular, it allows us to derive results 1) and 2) of section \ref{sec:newSQCD}.

\begin{figure}[!t]
\centering
\includegraphics[width=.3\textwidth]{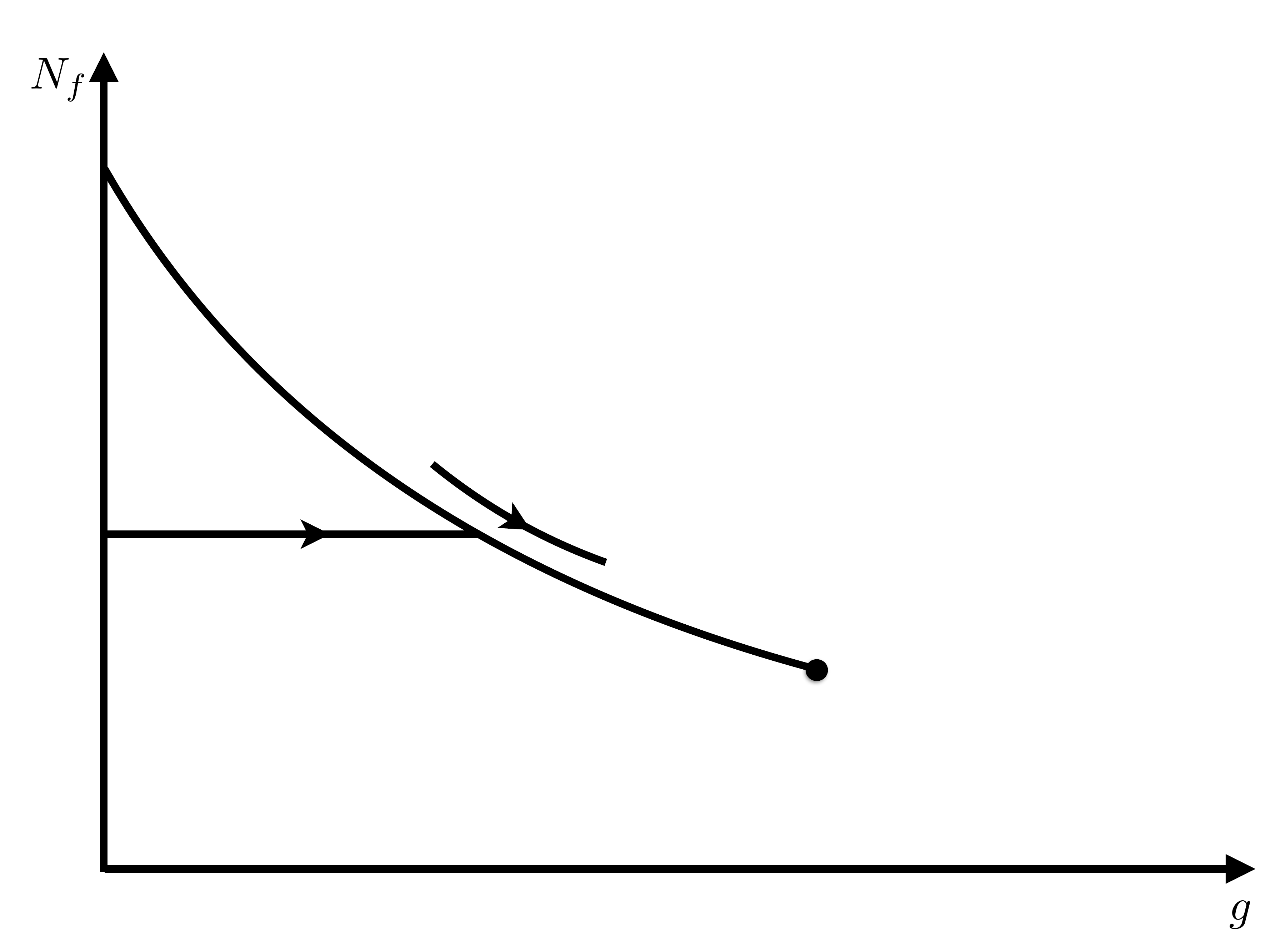}
\hfill
\caption{Flow I (horizontal line) from the UV ($g\!=\!0$) to the nontrivial IR fixed point for a theory with $N_f$ massless flavours in the conformal window. Flow II along the IR fixed point curve from a theory with $N_f$ massless flavours (UV) to one with $N_f'<N_f$ (IR).     } 
\label{fig:flows}
\end{figure}

Two types of UV to IR flows are of interest in this analysis, both were discussed in \cite{Anselmi:1997ys} and they are illustrated in Figure \ref{fig:flows}:

I. For $N_f$ fixed, the theory flows from the asymptotically free fixed point (UV) to the nontrivial IR fixed point, the horizontal line in Figure \ref{fig:flows}; we refer to this flow as UV${}_{AF}$.
The interpolating $a$-function $a(g(\mu))$, with renor\-malisation-scale dependent coupling $g(\mu )$, varies from its value $a_{UV}$ to  $a_{IR}$. We shall use the universality of the $a$-function to also derive constraints on the RG flow from a hypothetical strongly coupled UV fixed point to the weakly coupled IR fixed point; we refer to this flow as UV${}_{SC}$.  

II. One can devise a flow in the space of theories along the IR fixed point curve from a theory with $N_f$ massless flavours to one with $N_f-n$ massless flavours, and  $N_f-n\geqslant N_f^c$, so that both theories are in the same phase. This can be achieved by adding a mass deformation for $n$ flavours.
The interpolating $a$-function varies from $a_{UV}\!=\!a(N_f)$ to $a_{IR}\!=\!a(N_f-n)$.

In case I, from the IR Euler anomaly coefficient \cite{Anselmi:1997ys}\footnote{This result is valid to all orders in perturbation theory.} 
\be
\label{eq:aIR}
a_{IR}= \frac{3}{32}\left ( 2(N^2-1) +2N_f N (1-R)\left(1-3(1-R)^2\right)\right )\,,
\ee
 with $R\!=\!(2+\gamma_m^*)/3$, and $a_{IR}\!\rightarrow\!a_{UV}$ for $R\!\rightarrow\! 2/3$, one obtains for the flow  UV${}_{AF}$ \cite{Anselmi:1997ys}
\ba
\label{eq:aUVIR}
\hspace{-0.7cm}a_{UV}-a_{IR}&=& \frac{NN_f}{48}\gamma_m^{*2}\left (3-\gamma_m^*\right ) \nonumber\\
 &=&\frac{NN_f}{48}\left (1-\frac{3N}{N_f}\right )^2\left(2+\frac{3N}{N_f}\right)\, ,
\ea
where (\ref{eq:gamma_at_IRFP}) is used in the second line.  It vanishes at the upper edge, $N_f\!=\!3N$, and it
 satisfies $a_{UV}-a_{IR} >0$, $a_{IR}>0$, for $3N/2\leqslant N_f <  3N$ \footnote{ Corrections to $a_{IR}$ from a possible accidental symmetry due to the violation of the unitarity bound at $N_f\!=\!3N/2$ vanish \cite{Anselmi:1997ys}.}. 

To establish result 1) of section \ref{sec:newSQCD} we use the interpolating 
function $a(g(\mu ))$  for $N_f$ fixed, given by $a_{IR}$ in (\ref{eq:aIR}) for $\gamma_m^*\rightarrow\gamma_m(g(\mu ))$.
According to the $a$-theorem, $a(g(\mu ))$ is a strictly monotonic function of the scale $\mu$ and decreases from the UV to the IR, and it is stationary at a fixed point. Thus, away from the fixed point $\beta (g)\!\neq\!0$ and along the flows of type I with fixed $N_f$ 
\be
\label{eq:a_signs}
\frac{da}{d\log\mu} = \frac{\partial a}{\partial g}\beta (g) >0
\ee
implies that $\partial a/\partial g$ has the same sign as $\beta (g)$ with
\be
\label{eq:a_vs_g}
\frac{\partial{a}}{\partial g} 
= - \frac{NN_f}{16}\gamma_m\left (2-\gamma_m\right) \frac{\partial{\gamma_m}}{\partial g} \neq 0\,.
\ee 
For $\gamma_m<0$ and $\gamma_m>2$ ($0<\gamma_m <2$), it follows from (\ref{eq:a_vs_g}) that $\partial{\gamma_m}/{\partial g}\!\neq\!0$ and of the same sign (opposite sign) of ${\partial{a}}/{\partial g}$. Therefore, for $N_f$ fixed $\gamma_m$ must be a strictly monotonic function of $g$ away from fixed points. 
Importantly, this result applies to both flows, UV${}_{AF}$ ($\beta (g)<0$) and UV${}_{SC}$ ($\beta (g)>0$), given the universality of the interpolating $a$-function and (\ref{eq:a_vs_g}). This is result 1), and it will imply the incompatibility of SQCD with merging and result 2) in section \ref{sec:UV-IRmerging}.   
  
At the nontrivial fixed point, UV or IR, the flow of the $a$-function is stationary, $da/d\log\mu\!=\!0$, because $\beta (g)\!=\!0$, and 
(\ref{eq:a_vs_g}) does not constrain $\partial{\gamma_m}/{\partial g}$ -- unless one is able to prove that $\partial{a}/{\partial g}\!\neq\!0$ for any $g\!\neq\!0$. 

In case II, using (\ref{eq:aIR}) and  (\ref{eq:gamma_at_IRFP})  along the IR fixed point curve, one has
\be
\label{eq:DeltaNf}
a(N_f)-a(N_f-n)= \frac{9N^4}{16}\left ( \frac{1}{(N_f-n)^2} -      \frac{1}{N_f^2}\right ) > 0\,,
\ee
with $a(N_f)=(3N^2/16)(1-3N^2/N_f^2)$.  
In other words, the flow of $\gamma_m^*(N_f)$ implied by (\ref{eq:gamma_at_IRFP}), $d\gamma_m^*/dN_f >0$, guarantees that 
$da/dN_f=9N^4/8N_f^3$ is also positive, and results 3) and 4) can be re-interpreted as consequences of the $a$-theorem.   

Consistency of the $a$-theorem with result 5), through (\ref{eq:a_vs_g}), is obvious at this point, because (\ref{eq:a_vs_g}) does not constrain $\partial a/\partial g$ along the IR curve.
\subsection{Proof of the absence of merging in SQCD}
\label{sec:UV-IRmerging}
In this section we explore a specific mechanism that may lead to the occurence of the lower edge of a conformal window, guided  by the idea that such a mechanism is itself a powerful probe of the underlying theory. We consider the possibility that a nontrivial, i.e., interacting UV fixed point exists in the conformal window and merges with the IR fixed point at the lower edge.
The possibility of an additional, more strongly coupled UV fixed point in the QCD conformal window was put forward in \cite{Banks:1981nn}. The  merging of the UV-IR pair of fixed points at the lower edge \cite{Kaplan:2009kr,Gies:2005as} is  phenomenologically interesting, since it naturally leads to BKT/Miransky scaling \cite{PhysRevD.58.105017, Miransky:1996pd, Miransky1985,Berezinsky:1970fr, Berezinsky:1972, Kosterlitz:1973xp} and a ``walking'' gauge coupling just below the conformal window. 

Firstly, we establish a general result valid for SQCD and QCD:  A strictly monotonic $\gamma_G^*(N_f)$ and nonvanishing at the lower edge of the conformal window is incompatible with merging. Secondly, as an instructive exercise, we analyse merging in the context of SQCD and prove the incompatibility of the SQCD exact beta function with merging, by use of the $a$-theorem and result 1) of section \ref{sec:newSQCD}. 

\begin{figure}[!t]
\centering
\includegraphics[width=.4\textwidth]{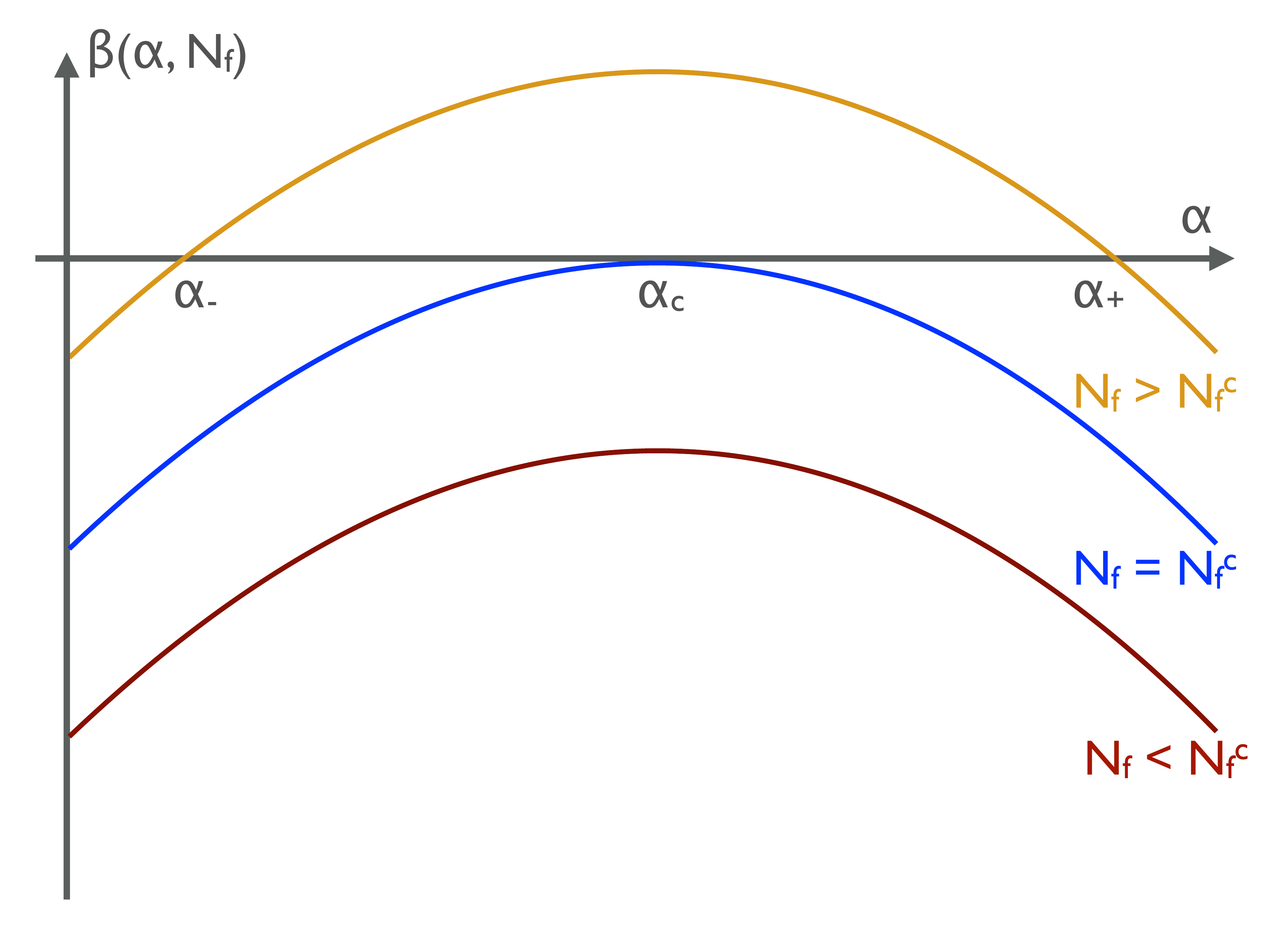}
\hfill
\caption{ The beta function $\beta (\alpha ,\, N_f)$ with $f(\alpha)\!=\!1$ in  (\ref{eqn:prototypeBeta}) for decreasing $N_f$, top to bottom: for $N_f>N_f^c$ there is a pair of fixed points at $\alpha_-$ (IR) and $\alpha_+$ (UV). They merge at $\alpha_c$ for $N_f\!=\!N_f^c$ and disappear for $N_f<N_f^c$. } 
\label{fig:beta-shapes-bkt}
\end{figure}

Close to $N_f^c$, the ansatz for the beta function that realises merging has the form \cite{Kaplan:2009kr} sketched in Figure \ref{fig:beta-shapes-bkt}:
\be
\beta(\alpha, \epsilon) = f(\alpha )\,\left [\epsilon - (\alpha -\alpha_c)^2\right ]\,,
\label{eqn:prototypeBeta}
\ee
where $\epsilon\!=\! (N_f-N_f^c)/N$, $\alpha$ is (a power of) a coupling, and $f(\alpha )$ is a strictly monotonic function of $\alpha$ \footnote{$f(\alpha )\!=\!1$ in \cite{Kaplan:2009kr}.}, nonzero on the interval $[\alpha_-,\alpha_+]$, with $\alpha_{\pm}\!=\! \alpha_c \pm \sqrt{\epsilon}$  the zeroes of $\beta (\alpha, \epsilon )$; $\alpha_\pm$ are distinct and real for $\epsilon>0$, $\alpha_+\!=\!\alpha_-\!=\!\alpha_c$ for $\epsilon\!=\!0$, and complex for $\epsilon <0$, thus leading to the disappearance of the conformal window. We note that the only effect of a strictly increasing (decreasing) $f(\alpha )$ in (\ref{eqn:prototypeBeta}) is to shift the maximum of the beta function, which occurs at $\alpha_c$ for $\epsilon\!=\!0$, to $\alpha_*>\alpha_c$ $(\alpha_*<\alpha_c)$ and $\alpha_-<\alpha_*<\alpha_+$ for $\epsilon >0$.   

At the lower edge, $\epsilon\!=\!0$,  
the beta function (\ref{eqn:prototypeBeta}) develops a local maximum at $\alpha_c$, thus $\beta'(\alpha_c, \epsilon\!=\!0)$ vanishes.
SQCD, like QCD, has one coupling, the gauge coupling, and the latter result, via 
(\ref{eq:gammaG_FP}) and $\alpha\!\sim\!g^2$, 
 implies that  $\gamma_G^*$ vanishes for $N_f\!=\!N_f^c$, though the theory is interacting. In other words, a nonvanishing $\gamma_G^*$ at the lower edge of SQCD, and QCD, is incompatible with merging. 

Besides, 
since $\gamma_G^*$ also vanishes at the upper edge, where the theory is IR free, 
and below the upper edge $\gamma_G(\alpha_\pm ,\epsilon >0 )\!=\!\beta'(\alpha_\pm, \epsilon >0)\!=\!\mp f(\alpha_\pm)\sqrt{\epsilon}$ is positive at $\alpha_-$ (IR) and negative at $\alpha_+$ (UV), then $\gamma_G^*(N_f)$ is non monotonic along the IR fixed point curve if merging occurs.

We now specialise to SQCD. The incompatibility of the SQCD exact beta function (\ref{eq:NSVZ}) with merging is a direct consequence of the $a$-theorem, through result 1) in section \ref{sec:newSQCD}, applied to the RG flow of the theory from a hypothetical strongly coupled UV fixed point to the weakly coupled IR fixed point. To prove it, we impose that (\ref{eq:NSVZ}) realises the merging form (\ref{eqn:prototypeBeta}) in the surroundings of the lower edge, $\epsilon\geqslant 0$ with $\epsilon \ll 1$. We equate 
(\ref{eqn:prototypeBeta}) with $\alpha\!=\!Ng^2$ to the SQCD beta function $\beta (\alpha)\!=\!2Ng\beta (g)$, with $\beta (g)$ in (\ref{eq:NSVZ}) \footnote{ This guarantees the correct $N, N_f$ counting for SQCD in the presence of merging.}.

Merging is realised for $f(\alpha)\!=\!\alpha^2/(8\pi^2 (1\!-\!\alpha/(8\pi^2) ) )$ strictly increasing on $[\alpha_-,\alpha_+]$, where as always $1\!-\!\alpha/8\pi^2 >0$, and 
\be
\label{eq:incomp}
-3 +(3/2+\epsilon )(1-\gamma_m (\alpha,\epsilon )) = \epsilon -(\alpha-\alpha_c)^2\,,
\ee 
where we used $N_f^c/N\!=\!3/2$ and  $N_f/N\!=\!3/2+\epsilon$. 
The condition (\ref{eq:incomp}) determines the RG flow of $\gamma_m$ on the interval $[\alpha_-,\alpha_+]$, for some $\epsilon \geqslant 0$, $\epsilon\ll 1$:
\be
\gamma_m(\alpha,\epsilon ) = \frac{-1 + (2/3)(\alpha-\alpha_c)^2}{1+2\epsilon /3}\, .
\ee
Thus $\gamma_m (\alpha_c, \epsilon )\!=\!-1/( 1+2\epsilon /3)$ is a minimum of $\gamma_m(\alpha,\epsilon )\!=\! \gamma_m (\alpha_c, \epsilon ) +(2/3) (\alpha -\alpha_c)^2/(1+2\epsilon /3)$ on the interval $[\alpha_-, \alpha_+]$, with $\alpha_\pm\!=\! \alpha_c\pm \sqrt{\epsilon}$. At the zeroes, $\gamma_m (\alpha_\pm , \epsilon )\!=\! -(1-2\epsilon /3)/(1+2\epsilon/3)$. 
Crucially, for any $\epsilon >0$, $\alpha_c$ does not correspond to a fixed point, nevertheless we have found that the RG flow of $\gamma_m$ is stationary at $\alpha_c$, if merging is realised. Equation (\ref{eq:a_vs_g}) then implies that the $a$-function itself is stationary at $\alpha_c$, away from a fixed point, thus violating the $a$-theorem.   
This establishes an important result, to all orders in perturbation theory: If the SQCD exact beta function satisfies the $a$-theorem, then it cannot realise merging. 

Even without the aid of an exact solution for the underlying theory, the non monotonicity of the scalar glueball anomalous dimension $\gamma_G^*(N_f)$ with merging along the IR fixed point curve seems at odds with the simple fact that interactions become stronger as $N_f$ becomes smaller along the IR fixed point curve, a feature implicit in the $a$-theorem. In fact, two-loop perturbation theory for QCD in section \ref{sec:pert_results}, as well as results 3), 4) and 5) for SQCD in section \ref{sec:newSQCD} are consistent with a nonvanishing $\gamma_G^*(N_f)$ everywhere below the upper edge, and monotonic along the IR curve; in SQCD, through (\ref{eq:NSVZderiv}), the latter properties hold for $\gamma_G^*(N_f)$ as well as the derivative of the mass anomalous dimension $\gamma_m^{\prime *}(N_f)$.

Finally, note that in the presence of merging the operator $\mbox{Tr}G^2$ would be irrelevant along the IR curve and relevant along the UV curve, marginal at the lower edge. 
Thus, plausibly, the UV fixed point curve would be a line of critical points in the conformal window, where a phase transition occurs in the continuum theory; this is a distinctive signature of merging. 

\subsection{Proof of result 2)}
\label{sec:result2}
Result 2) of section \ref{sec:newSQCD}, a result stronger than the incompatibility with merging,  follows straightforwardly from a similar line of reasoning. 
Consider the RG flow from a hypothetical nontrivial UV fixed point, with coupling $\alpha_{\scriptscriptstyle{UV}}\!\neq\!0$, to a nontrivial IR fixed point, with  coupling $\alpha_{\scriptscriptstyle{IR}}\!\neq\!0$, for $N,\,N_f$ fixed. 
If the beta function (\ref{eq:NSVZ}) is continuous and only vanishes at the fixed points $\beta(\alpha_{\scriptscriptstyle{IR,UV}})\!=\!0$, then two cases are possible:
\ba
\label{eq:2cases}
&&\hspace{-0.8cm} a) \,\mbox{If}~ 0<\alpha_{\scriptscriptstyle{IR}}<\alpha_{\scriptscriptstyle{UV}},~\beta(\alpha ) >0 ~\mbox{on}~(\alpha_{\scriptscriptstyle{IR}}, \alpha_{\scriptscriptstyle{UV}})  \nonumber\\
&&\hspace{-0.8cm} b)\,\mbox{If}~  0<\alpha_{\scriptscriptstyle{UV}}<\alpha_{\scriptscriptstyle{IR}},~  \beta (\alpha )<0~\mbox{on}~
(\alpha_{\scriptscriptstyle{UV}}, \alpha_{\scriptscriptstyle{IR}})\,.
\ea
We consider the first case, the top curve in Figure \ref{fig:beta-shapes-bkt}, and for convenience we write (\ref{eq:NSVZ}) as follows:
\ba
\label{eq:product}
\beta(\alpha)&=&f(\alpha)\,h(\alpha) \nonumber\\
f(\alpha)&=&\frac{\alpha^2}{8\pi^2 \left (1-\frac{\alpha}{8\pi^2}\right )} \nonumber\\
h(\alpha)&=&-3N+N_f -N_f\gamma_m(\alpha )\,.
\ea
Since both fixed points are at nonzero coupling, $f(\alpha )$ does not vanish on the closed interval $[\alpha_{\scriptscriptstyle{IR}},\alpha_{\scriptscriptstyle{UV}}]$ and it is continuous, with $1\!-\!\alpha/(8\pi^2) >0$.
Hence, the beta function vanishes only if its numerator vanishes, i.e., $\beta (\alpha_{\scriptscriptstyle{IR,UV}})\!=\!0$ only if $h (\alpha_{\scriptscriptstyle{IR,UV}})\!=\!0$, and the continuity of $\beta (\alpha )$ implies the continuity of $h(\alpha)$. 
Then $\beta (\alpha )$, continuous and vanishing only at the boundaries of $[\alpha_{\scriptscriptstyle{IR}},\alpha_{\scriptscriptstyle{UV}}]$, has a maximum at some  $\alpha_{\scriptscriptstyle{IR}}<\overline{\alpha} <\alpha_{\scriptscriptstyle{UV}}$, and $h(\alpha )$ also has a maximum at some  $\alpha_{\scriptscriptstyle{IR}}<\overline{\overline{\alpha}} <\alpha_{\scriptscriptstyle{UV}}$, and, by (\ref{eq:product}), $\gamma_m(\alpha )$ has an extremum at $\overline{\overline{\alpha}}$, away from a fixed point. 
Equation (\ref{eq:a_vs_g}) then implies  a stationary $a$-function away from a fixed point, hence the violation of the $a$-theorem. 

For the second case in (\ref{eq:2cases}), with $0<\alpha_{\scriptscriptstyle{UV}}<\alpha_{\scriptscriptstyle{IR}}$, 
the proof is  fully analogous, with the obvious exchanges of maxima and minima, IR and UV.

This proves that the $a$-theorem implies that the SQCD beta function 
does not admit more than one fixed point at nonzero coupling.
Hence, in the conformal window, $3N/2\leqslant N_f<3N$, the existence of the IR fixed point at nonzero coupling excludes an UV fixed point at nonzero coupling.

If one of the two fixed points occurs instead at zero coupling, the $a$-theorem can be satisfied, but not always. 
Consider the first case in (\ref{eq:2cases}), where now $\alpha_{\scriptscriptstyle{IR}}\!=\!0$ and $\alpha_{\scriptscriptstyle{UV}}>0$. This case could be realised above the conformal window, $N_f\geqslant 3N$, once asymptotic freedom is lost.
This time $f(\alpha_{\scriptscriptstyle{IR}})\!=\!0$ and $f(\alpha)$ is strictly positive and strictly increases on $(0, \alpha_{\scriptscriptstyle{UV}}]$, i.e., $f'(\alpha)\!=\!\alpha/(4\pi^2)(1-\alpha/(16\pi^2))/(1-\alpha/(8\pi^2))^2>0$,
so that $\beta (\alpha )$ has a maximum at some $0<\overline{\alpha} <\alpha_{\scriptscriptstyle{UV}}$ while 
$h(\alpha)$, and thus $\gamma_m(\alpha)$, are allowed to vary strictly monotonically on $(0, \alpha_{\scriptscriptstyle{UV}})$. Specifically, $h(\alpha)$ should vary from 
$h(\alpha_{\scriptscriptstyle{IR}})>0$ to $h(\alpha_{\scriptscriptstyle{UV}})\!=\!0$ and the $a$-theorem requires that it varies (decreases) strictly monotonically, i.e., $h'(\alpha)\!=\!(\beta(\alpha)/f(\alpha))'<0$ on $(0, \alpha_{\scriptscriptstyle{UV}})$, or equivalently
$\beta'(\alpha)/\beta(\alpha) < f'(\alpha)/f(\alpha)$ on $(0, \alpha_{\scriptscriptstyle{UV}})$ where $f, f'>0$ and $\beta >0$.

For $h(\alpha)$, and thus $\gamma_m(\alpha)$, strictly monotonic, and $\gamma_m\!\neq\!0,\,2$, equation (\ref{eq:a_vs_g}) then implies a strictly monotonic $a$-function on $(0, \alpha_{\scriptscriptstyle{UV}})$. However, 
the $a$-theorem through (\ref{eq:a_signs}) further requires that $\partial a/\partial\alpha$ has the same sign as $\beta(\alpha)$ away from a fixed point, hence $\partial a/\partial\alpha >0$ for $\beta(\alpha)>0$. 
Equation (\ref{eq:a_vs_g}) then implies the constraints: $\partial\gamma_m/\partial\alpha >0$ for $\gamma_m<0$ and $\gamma_m>2$, and $\partial\gamma_m/\partial\alpha <0$ for $0<\gamma_m<2$. 
Consider $\alpha$ in a neighbourhood of the origin $\alpha_{\scriptscriptstyle{IR}}\!=\!0$, with $\gamma_m(0)\!=\!0$. 
Then, $\partial\gamma_m/\partial\alpha >0$ if $\gamma_m(\alpha)>0$ and 
$\partial\gamma_m/\partial\alpha <0$ if $\gamma_m(\alpha)<0$. None of the latter solutions satisfies the $a$-theorem constraints above. This implies the absence of a nontrivial UV fixed point above the conformal window of SQCD, $N_f\geqslant 3N$ \footnote{This result was argued with different methods, using specific values of R-charges at the fixed points, in \cite{Intriligator:2015xxa}.}. 

The second case in (\ref{eq:2cases}), where now $\alpha_{\scriptscriptstyle{UV}}\!=\!0$ and $\alpha_{\scriptscriptstyle{IR}}>0$ is realised by the conformal window and can indeed be shown to be allowed by the $a$-theorem following a fully analogous proof. 
It is worth to note that none of these proofs make use of specific assignments of R-charges at fixed points, nor of their uniqueness.
\section{Large-N QCD in the Veneziano limit}
\label{sec:largeN_results}
We now investigate to what extent the results obtained in SQCD remain valid in the massless Veneziano limit  ($N_f, N$ $\rightarrow\infty$, $N_f/N\!=\!\mbox{const}$)
of large-$N$ QCD, for which an exact beta function has been proposed \cite{Bochicchio:2013aha}, as a generalisation of the large-N Yang-Mills exact beta function derived on the basis of the loop equations for certain quasi-BPS Wilson loops \cite{Bochicchio:2008vt}. 
This beta function remarkably manifests salient analogies with the exact NSVZ beta function in (\ref{eq:NSVZ}), with one crucial difference. From inspection of the beta function  for given $N$, $N_f$ \footnote{Ref. \cite{Bochicchio:2013aha} writes (\ref{eq:betaNQCD}) in terms of the 't Hooft coupling $g_c\!=\!\sqrt{N}g$.} \cite{Bochicchio:2013aha,Bochicchio:2008vt}
\ba
\label{eq:betaNQCD}
&&\hspace{-1.3cm}\beta(g) =\frac{\partial g}{\partial\log\mu} = \\
&&\hspace{-0.8cm}-\frac{g^3}{16\pi^2}
\frac{ (4\pi )^2\beta_0 -N \left ({\partial\log Z}/{\partial\log\mu}\right ) +N_f \gamma_m(g)    } {1-N\left ( g^2/4\pi^2\right )}\, ,
\nonumber
\ea
with $\beta_0$ in (\ref{eq:b12}), the anomalous dimension factor 
\ba
\label{eq:gamma0Z}
\frac{{\partial\log Z}}{{\partial\log\mu}}\!&=&\!2\gamma_0\left ( Ng^2+\ldots \right )\nonumber\\
\gamma_0\!&=&\!\frac{5}{3(4\pi )^2}\left (1-\frac{2N_f}{5N}\right )
\ea
and the fermion mass anomalous dimension 
\be
\gamma_m(g) = -\frac{9}{3(4\pi)^2}\frac{N^2-1}{N} g^2+\ldots\, ,
\ee
both starting at order $Ng^2$, and comparing with (\ref{eq:NSVZ}), one concludes that the absence of supersymmetry generates the new anoma\-lous dimension contribution $\partial\log Z/\partial\log\mu$ in the beta function of QCD; its structure  is otherwise identical to (\ref{eq:NSVZ}). 
Equation (\ref{eq:betaNQCD}) is exact in the large-N limit, i.e., to leading order in the $1/N$ expansion, and it is exact to all orders in the $O(1)$ ratio $N_f/N$ in the Veneziano limit. Indeed, one can verify that 
its weak coupling expansion reproduces the universal part of the perturbative beta function, i.e., the two-loop order, up to the last
contribution to the two-loop coefficient $\beta_1$ in (\ref{eq:b12}), which is $1/N^2$ suppressed with respect to the leading contribution
\cite{Bochicchio:2008vt,Bochicchio:2013aha}.

Another very interesting result \cite{Bochicchio:2013aha} is the determination of the lower edge of the conformal window, within the local approximation of the glueball effective action valid in the confining phase. The lower edge occurs at  $N_f/N\!=\!5/2$, the value for which $\gamma_0$ in (\ref{eq:gamma0Z}) changes sign. In fact, $\gamma_0$ also enters the glueball kinetic term, and its change of sign signals a phase transition from confinement to a phase with $\langle{\mbox{Tr}}(G^2)\rangle\!=\!0$, the conformal Coulomb phase. 

Then, for $N_f/N\!=\!5/2$, barring the occurrence of a cusp singularity and noting that  ${\partial\log Z}/{\partial\log\mu}\!=\!0$ \footnote{The anomalous dimension term has an exact expression with overall coefficient $\gamma_0$ in terms of the Wilsonian coupling \cite{Bochicchio:2008vt,Bochicchio:2013aha}.}, the beta function (\ref{eq:betaNQCD}) vanishes for $\gamma_m\!=\!-4/5$, a renormalisation-scheme independent result. As anticipated in section \ref{sec:pert_results}, this result suggests that the singularity of the QCD two-loop beta function at the lower edge, i.e., $g_*\rightarrow\infty$ for $\beta_1=0$, is indeed an artefact of the truncated perturbative expansion.

We determine $\gamma_G^*(N_f)$  using (\ref{eq:betaNQCD}):
\be
\label{eq:largeNderiv}
\gamma_G^*(N_f) =
-\frac{g_*^3}{16\pi^2}\,\frac{ -N({\partial\log Z}/{\partial\log\mu})^{\prime *} (N_f)+N_f\gamma_m^{\prime *}(N_f)}{ 1-Ng_*^2/(4\pi^2)}\, ,
\ee
where from (\ref{eq:gamma0Z}) the derivatives with respect to $g$ are
\begin{eqnarray}
\label{eq:derZgamma}
({\partial\log Z}/{\partial\log\mu})' &=& 2 \gamma_0 \left (2Ng +\ldots \right ) \nonumber\\
\gamma'_m&=& -\frac{6}{(4\pi )^2} \frac{N^2-1}{N}\, g+ \ldots 
\end{eqnarray}
At the lower edge the derivative $({\partial\log Z}/{\partial\log\mu})^{\prime *} (N_f^c)$ vanishes exactly since $\gamma_0\!=\!0$. 
Therefore, at the lower edge (\ref{eq:largeNderiv}) reduces to 
\be
\label{eq:gamma_G_LE}
\gamma_G^*(N_f^c)= -\frac{g_*^3}{16\pi^2}\,\frac{ N_f\gamma_m^{\prime *}(N_f^c)   }{ 1-Ng_*^2/(4\pi^2)}\, ,
\ee
which is the main result of this section, 
a relation between $\gamma_G^*$ and $\gamma_m^{\prime *}$ entirely analogous  to SQCD. 
Like (\ref{eq:betaNQCD}), (\ref{eq:gamma_G_LE}) is exact in the large-N limit and to all orders in the $O(1)$ ratio $N_f/N$ in the Veneziano limit.
Barring the occurrence of a cusp singularity, it suggests that the singular behaviour of two-loop perturbation theory, $\gamma_G^*\rightarrow\infty$ for 
$\beta_1\!=\!0$,  is an artefact of the perturbative expansion. 

In full analogy with SQCD, equation (\ref{eq:gamma_G_LE}) implies that $\gamma_G^*(N_f^c)$ is strictly positive, if $\gamma_m^{\prime *}(N_f^c)<0$. 
The latter condition is at least verified in (\ref{eq:derZgamma}) to leading order in perturbation theory \footnote{Note, however, that $g_*$ comes from the cancellation of a priori infinitely many terms in the expansion in $g$.}, and there are no physical constraints that force $\gamma_m^{\prime *}$ to vanish at the lower edge, analogously to SQCD. 

A nonvanishing $\gamma_G^*$ at the lower edge would then exclude merging, according to section \ref{sec:UV-IRmerging}, and it would lead to the following description. 
 A phase transition occurs at the lower edge, and $\gamma_G$ develops a finite discontinuity:
$\gamma_G^*(N_f)$ is given by (\ref{eq:gamma_G_LE}) and is positive for $N_f/N\!=\!5/2$, while in the absence of a fixed point $\gamma_G$ is given by (\ref{eq:gammaG}) and is negative for $N_f/N<5/2$, the confining phase, without vanishing -- note that $\gamma_0$ in (\ref{eq:gamma0Z}) and $({\partial\log Z}/{\partial\log\mu})'$ in (\ref{eq:derZgamma}) no longer vanish below the lower edge.
\subsection{QCD and the $a$-theorem}
\label{sec:QCDatheorem}
What about the $a$-theorem and its constraints on QCD or any of its limits? 
The $a$-theorem, as proved in \cite{Komargodski:2011vj}, would imply the existence of a proper UV to IR interpolating $a$-function for a vast class of four-dimensional field theories where a unitary S matrix exists, thus including SQCD, as well as QCD.
On the other hand, one can construct an $a$-function and study $a_{UV}$ and $a_{IR}$ only in a limited set of examples. 
In SQCD, supersymmetry and the exact anomaly-free R symmetry at the fixed point are the key properties that allow the explicit construction of the interpolating $a$-function  discussed in section \ref{sec:atheorem}. Most importantly, they allow us to show how the $a$-function evolution directly constrains the ultraviolet to infrared flow of the mass anomalous dimension and its derivatives. 

In QCD, some results are also available. Cardy's conjectured $a$-function, which coincides by construction with the Euler anomaly coefficient $a_{UV}$ ($a_{IR}$) at the UV (IR) CFTs, has been shown to satisfy $a_{UV}-a_{IR}>0$ in the confined and chirally broken phase of QCD, when its infrared realisation is assumed to have $N_f^2-1$ massless Goldstone bosons that are free in the long distance limit \cite{Cardy:1988cwa}. This result is not based on perturbation theory.
Close to the upper edge of the QCD conformal window, in the large-N limit and with $\epsilon\!=\!11/2-N_f/N\ll 1$, two-loop perturbation theory verifies the $a$-theorem, i.e.,  $a_{UV}-a_{IR}>0$ and of order $N^2\epsilon^2$ \cite{Jack:1990eb}.
The $a$-theorem in the context of the massless Veneziano limit of large-N QCD deserves further investigation. 
\subsection{QCD and merging}
\label{sec:QCDmerging}
Though the  validity of the Veneziano limit of large-N QCD exact beta function (\ref{eq:betaNQCD}) in the deconfined phase above the lower edge ($N_f>N_f^c$) has not yet been demonstrated, it is instructive to investigate its compatibility with the merging hypothesis. 
One can repeat the exercise done for SQCD in section \ref{sec:UV-IRmerging}. 
For the beta function $\beta (\alpha)\!=\!2Ng\beta (g)$, with $\beta (g)$ in (\ref{eq:betaNQCD}) and $f(\alpha )=\alpha^2/(8\pi^2 (1-\alpha/(4\pi^2) ) )$, the merging condition analogous to (\ref{eq:incomp}) reads:
\be
-2 +\frac{2}{3}\epsilon + \frac{\partial\log Z}{\partial\log\mu} -  \left (\frac{5}{2}+\epsilon\right )\gamma_m =\epsilon -(\alpha -\alpha_c)^2\,,
\ee 
where, from (\ref{eq:gamma0Z}), $\partial\log Z/\partial\log\mu = -(4/3(4\pi)^2)\epsilon (\alpha +\ldots )$ is of order $\epsilon$.
Thus, for $\epsilon\!=\!0$, $\gamma_m (\alpha , \epsilon\!=\!0)= -4/5 + (2/5)(\alpha -\alpha_c)^2$ has a minimum at $\alpha_c$.
This time, for $\epsilon >0$, we can at least conclude that the function 
\be
\label{eq:minQCD}
\left (1+\frac{2}{5}\epsilon\right ) \gamma_m (\alpha ,\epsilon ) -\frac{2}{5}\frac{\partial\log Z}{\partial\log\mu} (\alpha , \epsilon )
= -\frac{4}{5} -\frac{2}{15}\epsilon +\frac{2}{5}(\alpha -\alpha_c)^2
\ee
has a minimum at $\alpha_c$, away from a fixed point. Thus, differently from SQCD, contributions from $\partial\log Z/\partial\log\mu$ enter the merging condition to order $\epsilon$ as in (\ref{eq:minQCD}). We may expect that, in full analogy with SQCD, it is now the function in the lhs of (\ref{eq:minQCD}) that enters the interpolating $a$-function for the Veneziano limit of large-N QCD, so that a relation analogous to (\ref{eq:a_vs_g}) would again lead to the incompatibility of the Veneziano limit of large-N QCD with merging; we defer this analysis to future work. 
\subsection{Vanishing $\gamma_G$ and the free theory}
\label{sec:free}
Results 3) to 5) in section  \ref{sec:newSQCD} for SQCD, and QCD two-loop perturbation theory suggest that $\gamma_G$ does not vanish along the IR fixed point curve below the upper edge of the conformal window, including its lower edge.
However, even with the aid of an exact relation between $\gamma_G$ and $\gamma_m'$ at the fixed point, (\ref{eq:NSVZderiv}) for SQCD and (\ref{eq:gamma_G_LE})  for the Veneziano limit of large-N QCD, we could not exactly constrain  their fixed-point value at the lower edge.
Oppositely, we have shown that merging forces $\gamma_G^*(N_f)$ to vanish at the lower edge, and vary non monotonically with $N_f$.  

In this section we limit ourselves to note the following: Proving that, in d=4,  $\gamma_G\!=\!0$ at a fixed point implies a free CFT would directly guarantee  that $\gamma_G$ cannot vanish at the fixed point of  the lower edge of the conformal window, where the theory is interacting, thus excluding merging in QCD.  

A proof  would amount to show that a theory in d=4 with a scalar operator of scale dimension $\Delta\!=\!4$ ($\mbox{Tr}G^2$) is free. 
The conformal partial wave expansion has been fully worked out  for scalar fields \cite{Dolan:2011dv,Dolan:2003hv,Dolan:2000ut}, but  not many exact results are available in d=4. It has been shown that theories involving a scalar of dimension $\Delta\!=\!2$ are free \cite{Nikolov:2005yu,Bakalov:2007dc},
and that 
theories with an infinite number of conserved higher spin currents $(s>2)$ are free in d=3 \cite{Maldacena:2011jn}. The latter proof has been partially extended to d=4 \cite{Stanev:2012nq}.
One step forward would be to verify the agreement of the four point function of ${\mbox{Tr}}G^2$ in the $\gamma_G\!=\!0$ limit with the four point function of the same scalar operator in a theory with free Abelian vector fields derived in, e.g., \cite{Dolan:2000ut}. 
\subsection{Merging with multiple couplings}
\label{sec:merge_multiple}
It has been conjectured \cite{Kaplan:2009kr,Gies:2005as} that the description of strongly coupled QCD in the conformal window may involve, in addition to the gauge coupling, one or more effective couplings $c_i$, associated to effective composite operators $O_i$, e.g., a four-fermion operator whose coupling's beta function develops a pair of nontrivial IR and UV zeroes that realise merging at the lower edge as in Figure \ref{fig:beta-shapes-bkt}. In this scenario,
the gauge coupling beta function has only a nontrivial IR zero, but it could also develop an additional nontrivial UV zero. 

We should immediately realise that this description is simply excluded in QCD, as it is in SQCD, and, a fortiori in the Veneziano limit of large-N QCD description of the physics at the lower edge in section \ref{sec:largeN_results}, which is in terms of the gauge coupling only. In other words, even if we could find some additional composite operator that provides a correct effective description of QCD or the Veneziano limit of large-N QCD in some energy range, its coupling is fully determined by the gauge coupling, so that the RG flow of the theory is uniquely dictated by the gauge coupling beta function, the one in (\ref{eq:betaNQCD}) for the Veneziano limit of large-N QCD. 

The remainder of this section is a side note that discusses, consistently with the previous conclusion, how the IR fixed point of the QCD conformal window can no longer be recovered when  a generic four-fermion operator is added. Such an addition generally leads to a gauge-NJL model, and, for our argument, it is sufficient to consider a scalar four-fermion operator only. Then, bosonisation via an auxiliary scalar field of the gauge-NJL model leads to a gauge-Yukawa theory as its low-energy realisation. The RG flow from the UV to the IR renders the auxiliary scalar field dynamical and renormalises the scalar mass, the Yukawa coupling and the four-scalar interaction.
We are familiar with the RG flow from an asymptotically free UV to the IR.
Schematically, starting at $\Lambda$ with an irrelevant interaction $G(\bar{\psi}\psi )^2$, with $G\sim \Lambda^{-2}$, after integrating out one-fermion loops with momenta $\mu\leqslant p\leqslant \Lambda$ and rescaling to a canonically normalised scalar field, one has at the scale $\mu$
\be
\label{eq:NJL}
m_\phi^2(\mu ) =\frac{M_\phi^2(\mu )}{Z_\phi (\mu )}~~~a_y(\mu )=\frac{1}{Z_\phi (\mu )}~~~\lambda (\mu )=\frac{\lambda_0(\mu )}{Z_\phi^2(\mu )}
\ee
for the squared scalar mass, the squared Yukawa coupling and the four-scalar coupling, respectively, and $Z_\phi$ is the scalar wave function renormalisation. $\lambda_0$ and $Z_\phi$ have logarithmic running $\log (\Lambda^2/\mu^2)$, while $M_\phi^2 (\mu ) = M_\phi^2(\Lambda ) - (\Lambda^2 -\mu^2)$ has a quadratic running \footnote{A fine-tuning of $M_\phi^2(\Lambda ) -\Lambda^2$ in (\ref{eq:NJL}) is required by construction in the gauge-NJL model where a hard cutoff regularisation is used.}, with boundary conditions
\be
M_\phi^2(\Lambda )=1/G\sim \Lambda^2~~~\lambda_0(\Lambda ) =0~~~Z_\phi(\Lambda )=0\, .
\ee
One cannot recover
the original QCD IR fixed point in the conformal window, because the scalar cannot decouple from the IR spectrum. Firstly, note that the RG flow from a hypothetical strongly coupled UV fixed point, for $N_f$ fixed, must lead to the same IR fixed point as 
the RG flow from the asymptotically free UV -- these flows can be pictured in Figure \ref{fig:flows} as perturbations in a multiple-coupling space on the right side or the left side of the IR fixed point curve,  respectively.  
Secondly, the latter RG flow is often used to ``mimic'' the spontaneous breaking of chiral symmetry in QCD, at some scale $\mu <\Lambda$. This is the first one of two possible solutions in (\ref{eq:NJL}): i) The squared scalar mass $m_\phi^2$ becomes negative thus inducing spontaneous symmetry breaking, and the mass of the scalar fluctuations are proportional to the dynamically generated fermion mass, but it is not QCD, or ii) the scalar becomes free in the IR, or, provided its mass vanishes, a nontrivial IR fixed point can develop. 
The latter realises a massless gauge-Yukawa model with an IR fixed point, not QCD, and the same IR fixed point must be reached by the RG flow from the hypothetical strongly coupled UV fixed point where a four-fermion operator can eventually be marginal or relevant. The decoupling of the scalar field from the IR spectrum advocated in \cite{Kaplan:2009kr} is equivalent to taking $\Lambda\to\infty$ and remove the four-fermion operator at all scales. 

A perturbative analysis of the massless and chirally symmetric gauge-Yukawa theory with $N_f$ fundamental fer\-mi\-ons allows us to better understand how the QCD conformal window is modified, having clarified that is no longer QCD.
QCD symmetries are preserved when all $N_f$ fermions have degenerate Yukawa coupling to the appropriate combination of scalar and pseudoscalar fields. This is model C in \cite{Kaplan:2009kr}, where it is shown, consistently with the more general perturbative analysis in \cite{Antipin:2012kc}, that the theory in the Veneziano limit has no conformal window at the 2-1-1 (gauge-Yukawa-scalar) loop order; Yukawa interactions push the IR fixed point towards stronger coupling until the conformal window disappears. 
This also means that any nontrivial zero, both IR and UV, that could be generated at this or higher orders in perturbation theory for $N_f<N_f^{AF}$ 
has anyway no resemblance of the IR fixed point of the QCD conformal window. 
\section{Final remarks}
\label{sec:conc}
In this letter we have shown that the exact beta function of SQCD entails an exact relation between the anomalous dimension $\gamma_G$ of the scalar glueball operator and the derivative of the mass anomalous dimension  $\gamma_m$ at the IR fixed point in the conformal window and that, remarkably, the recently proposed exact beta function for the massless Veneziano limit of large-N QCD entails a fully analogous relation at the lower edge of the conformal window. We can view this relation as one way in which the gauge sector and the matter sector are intertwined in QCD.

The $a$-theorem has then allowed us to prove the incompatibility of the SQCD exact beta function with the merging of fixed points to all orders in perturbation theory, through constraints on the RG flow of the theory away from fixed points. The analogies with the massless Veneziano limit of large-N QCD then allowed us to suggest the way in which the same incompatibility may extend to QCD as a consequence of the $a$-theorem. By the same means we have also determined the general conditions under which the SQCD exact beta function satisfies the $a$-theorem, and, as a result, we have excluded the existence of more than one fixed point at nonzero coupling as well as
a nontrivial UV fixed point in the IR free theory above the conformal window.

We have shown that $\gamma_G$  carries information about the nature of the lower edge of the conformal window, $N_f\!=\!N_f^c$: 
A nonvanishing $\gamma_G$ at the lower edge of the QCD conformal window would exclude the merging of fixed points.  
At the same time, we have shown that SQCD in the Vene\-zia\-no limit and QCD two-loop perturbation theory are indeed consistent with a strictly positive and monotonically increasing $\gamma_G$ at the IR fixed point as $N_f$ decreases below the upper edge of the conformal window. 

It is worth noting that the prediction of the lower edge at $N_f/N\!=\!5/2$
 in the Veneziano limit of large-N QCD \cite{Bochicchio:2013aha,Bochicchio:2008vt} is in nice agreement with the recently determined bound on the lower edge $6 < N_f^c < 8$ for the $SU(3)$ theory \cite{daSilva:2015vna} based on a lattice QCD study.

We have also observed that a multiple-coupling merging, arising from the hypothesis that strongly coupled QCD may require additional 
composite operators in its description, is, by construction, incompatible with QCD, as it is with SQCD, and, a fortiori with the Veneziano limit of large-N QCD description of the lower edge of the conformal window.  

In light of this analysis, the combined nonperturbative determination of $\gamma_G$ and $\gamma_m$ along the IR fixed point curve in the conformal window, with lattice and/or conformal bootstrap techniques, would certainly be a useful test for QCD, able to unambiguously determine the mechanism in place for the onset of conformality.  
\section*{Acknowledgements}
We thank M. Bochicchio for many valuable comments and discussions. We also thank D. Anselmi, R. Crewther, L. Tunstall and R. Schrock for useful correspondence. 
\vspace{0.5cm}
\bibliography{mybibfile}

\end{document}